\documentclass{article}
\usepackage{spconf,amsmath,graphicx,hyperref,amssymb,mathtools}
\usepackage{cite}
\usepackage{booktabs}
\usepackage{tabularx}
\usepackage{array}
\usepackage{enumitem}
\newcommand{\accunit}{\textit{per piece macro (\%)}}
\newcommand{\smalltable}{\setlength{\tabcolsep}{4pt}\renewcommand{\arraystretch}{1.1}}

\title{BACHI: Boundary-Aware Symbolic Chord Recognition Through \\ Masked Iterative Decoding on Pop and Classical Music}
\name{Mingyang Yao, Ke Chen, Shlomo Dubnov, Taylor Berg-Kirkpatrick}
\address{
   University of California San Diego, USA
}
\begin{document}
\maketitle
\ninept

\begin{abstract}
Automatic chord recognition (ACR) via deep learning models has gradually achieved promising recognition accuracy, yet two key challenges remain. First, prior work has primarily focused on audio-domain ACR, while symbolic music (e.g., score) ACR has received limited attention due to data scarcity. Second, existing methods still overlook strategies that are aligned with human music analytical practices. To address these challenges, we make two contributions: (1) we introduce POP909-CL, an enhanced version of POP909 dataset with tempo-aligned content and human-corrected labels of chords, beats, keys, and time signatures; and (2) We propose BACHI, a symbolic chord recognition model that decomposes the task into different decision steps, namely boundary detection and iterative ranking of chord root, quality, and bass (inversion). This mechanism mirrors the human ear-training practices. Experiments demonstrate that BACHI achieves state-of-the-art chord recognition performance on both classical and pop music benchmarks, with ablation studies validating the effectiveness of each module.
\end{abstract}


\section{Introduction}
\label{sec:intro}
Automatic chord recognition (ACR) is a fundamental task in music information retrieval that aims to annotate music with chord labels over time. Recent breakthroughs in machine learning have shown that ACR models with deep neural networks outperform conventional approaches. Chord recognition also underpins a wide range of downstream applications, including harmonic analysis, annotation for controllable music generation, and music education.

Despite recent promising progress, symbolic chord recognition remains underexplored compared to its development in the audio domain, largely due to data scarcity. Audio-based ACR models benefit from well-established benchmarks in pop music, such as USPOP~\cite{USPOP}, RWC-Popular~\cite{RWC}, Billboard~\cite{Billboard}, Isophonics~\cite{Isophonics}, and CASD~\cite{CASD}. Although benchmarks for classical music are fewer, models trained on pop music have captured many learnable patterns and demonstrated broad applications in music analysis and music generation tasks. 

In contrast, symbolic ACR faces the opposite situation and a more severe imbalance. Very few datasets provide accurate chord annotations except for Hooktheory~\cite{hooktheory}, but it only includes chord labels and melody without full arrangements or textures, limiting its usability for ACR. For classical music, most chord-annotated symbolic datasets derive from the Roman Numeral Harmonic Analysis task. Among these, When-in-Rome~\cite{when-in-rome} and the DCML corpus~\cite{dcml} provide the largest collections (about 1700 works in total), while others, such as the Beethoven Sonata corpus~\cite{Beethoven_Sonatas}, TAVERN~\cite{TAVERN}, and BPS~\cite{BPSFH} contain fewer than 100 works each. However, these corpora often include duplicated pieces or short excerpts, and existing approaches could only rely on subsets or combined chunks for ACR training. For example, ChordGNN~\cite{chordgnn} and AugmentedNet~\cite{augmentednet} use 300+ works (1400+ segments), while Harmony Transformer~\cite{ht, htv2} uses fewer than 100 works, since it outputs Roman numeral analyses rather than chord labels alone.

Another major challenge of symbolic ACR lies in methodology. Prior approaches have adopted diverse strategies to improve performance. The rule-based method~\cite{rule-based} aggregates notes within windows to infer chords but often fails with complex harmonic progressions due to their limited reasoning capability. AugmentedNet~\cite{augmentednet} and ChordGNN~\cite{chordgnn} model structure note interactions using either convolutional or graph neural networks to strengthen the note–chord correlation in recognition. Harmony Transformer~\cite{ht, htv2} focuses more on the Roman numeral harmonic analysis task and introduces transformer architectures with global-context modeling, bridging long-range note dependencies. However, most approaches pay little attention to aligning the chord recognition process with the human annotation process, which is a potential factor that may be crucial for improving accuracy. Some audio-based ACR models have explored this idea. For exampe, ~\cite{audio_decompose} decomposes chord targets into constituent elements (root, triad, 7th) to guide recognition. Inspired by this, transferring domain knowledge from human annotation logic offers a promising direction for improving performance.

\begin{figure*}[t]
    \centering
    \vspace{-5mm}
    \includegraphics[width=\linewidth]{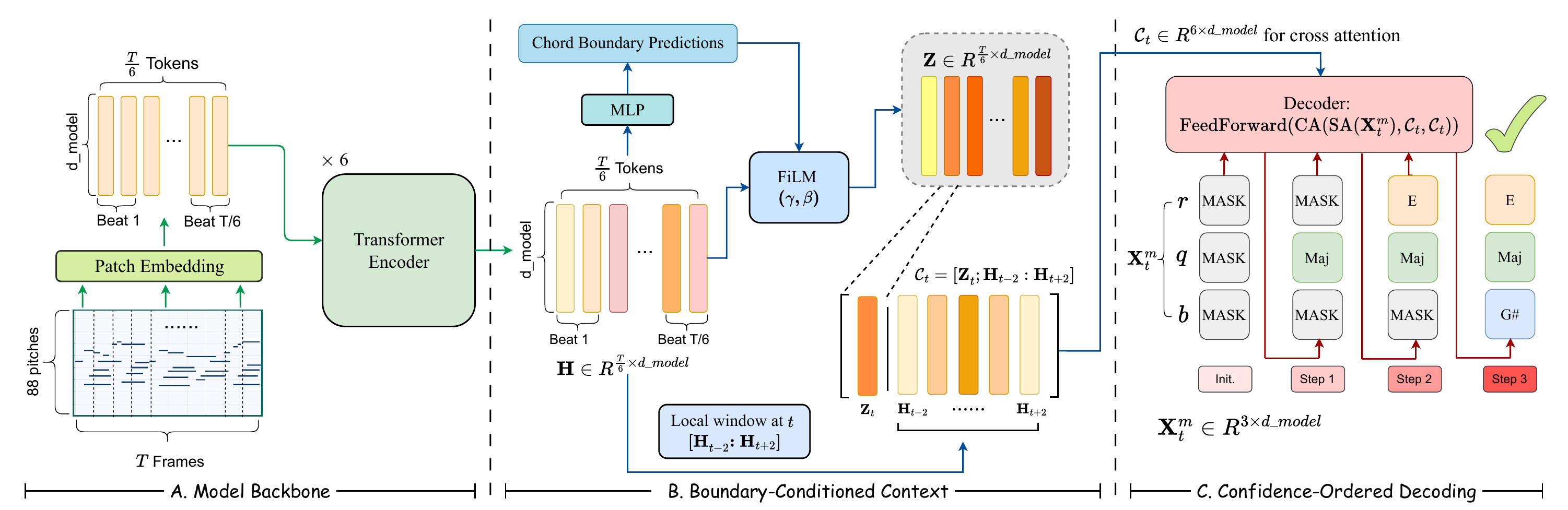}
    \caption{The model architecture and inference mechanism of BACHI, from the model backbone (left and middle), boundary detection and conditioning (middle), and iterative decoding (right).}
    \vspace{-0.2cm}
    \label{fig:BACHI}
\end{figure*}

In this paper, we address the challenges of data scarcity and methodology in symbolic chord recognition. First, we propose POP909-CL, an enhanced version of the POP909 dataset~\cite{pop909} with human-corrected labels for chords, beats, keys, and time signatures. This resource supports not only symbolic chord recognition but also broader symbolic MIR tasks. Second, we propose BACHI, a \textbf{B}oundary-\textbf{A}ware symbolic \textbf{CH}ord recognition model with masked \textbf{I}terative decoding. It operates on beat-synchronous MIDI tokens and integrates two key components: (1) a supervised boundary detection module that predicts chord-change likelihoods and modulates encoder states via feature-wise linear modulation~\cite{film} (FiLM); and (2) a transformer decoder~\cite{transformer} that iteratively predicts chord root, quality, and bass in confidence order. This design mirrors human sight-singing and ear-training practices, where chord perception emerges progressively through cues whose order depends on different contexts. Our contributions are three-fold:
\begin{itemize}[leftmargin=*]
    \item We proposed POP909-CL, an enhanced version of POP909 with human-corrected labels, as a reliable resource for MIR research.
    \item We propose BACHI, a boundary-aware symbolic chord recognition model that incorporates cues and decision processes inspired by human ear-training to improve recognition accuracy.
    \item Objective evaluations demonstrate that BACHI achieves state-of-the-art performance on both classical and pop music benchmarks, with ablation studies validating the effectiveness of each module.
\end{itemize}

\section{Method}\label{sec:method}
Figure~\ref{fig:BACHI} illustrates the model architecture and inference mechanism of BACHI. In the following subsections, we will introduce its input and output specifications and the different recognition stages.

\vspace{-0.2cm}
\subsection{Input and Output}\label{subsec:data}
For input, BACHI encodes symbolic musical scores as piano rolls $\mathbf{P} \in \{0,1\}^{T \times D}$, where $D=88$ is the dimensionality of pitch classes and the temporal resolution for $T$ is 12 frames per beat, keeping triplet and dotted 16th notes precise. We employ a patch embedding module to convert piano rolls into continuous latent tokens. Specifically, the patch embedding module contains a 1D-CNN layer with a kernel size 6 to map the pitch class channel $D$ to the hidden dimension of the transformer $d\_model=512$, and reduce the temporal dimension from $T$ to $T/6$. It is followed with a GLU activation layer to create a more normalized and compact representation. 

For output and groundtruth labels, we decompose each chord label into three elements: root, quality, and bass. Bass is referred to the bass note for chord inversion (e.g., C/G is a C major chord inversion with C root, major quality, and G bass). The label sequence is processed with the same temporal resolution as input ($T/6$).

\subsection{Two-stage Chord Recognition}\label{subsec:hierarchical}
After processing the input and output sequences, we adopt a two-stage chord recognition framework: (1) boundary detection, and (2) iterative chord element decoding using confidence ranking with masked transformers.

\vspace{0.2cm}
\noindent\textbf{Boundary Detection}\; The input music sequence is fed into six transformer encoder blocks to produce the hidden state sequence output $\mathbf{H}$. An MLP layer is then applied to predict the chord boundary sequence $\mathbf{e}$, where boundary labels are obtained as binarized labels derived from the chord labels. The predicted boundaries are further embedded as an additional condition for the subsequent chord recognition step via feature-wise linear modulation~\cite{film} (FiLM):
\begin{align}
[\boldsymbol{\gamma}_t,\boldsymbol{\beta}_t] &=  [\mathrm{MLP_{\gamma_t}}\big(\mathrm{LN ([\mathbf{H}_t;\mathbf{e}_t])\big)},\mathrm{MLP_{\beta_t}}\big(\mathrm{LN ([\mathbf{H}_t;\mathbf{e}_t])\big)}]\\
\mathbf{Z}_t &= \mathrm{LN}(\mathbf{H}_t)\odot (1{+}\boldsymbol{\gamma}_t) + \boldsymbol{\beta}_t.
\end{align}
Where $\gamma$ and $\beta$ denote the scale and the bias from FiLM via two MLP layers and LayerNorm~\cite{ln}. $\mathbf{Z}$ denotes the FiLM-conditioned latent representation from $\mathbf{H}$ and boundary predictions $\mathbf{e}$. 

\vspace{0.2cm}
\noindent\textbf{Iterative Decoding by Confidence} 
The framewise representation $\mathbf{Z}_t$ is combined with a local context window $r=2$ to form a new sequence $\mathcal{C}_t$=$ [\mathbf{Z}_t, \mathbf{H}_{t-2:t+2}] \in R^{6\times d\_model}$, which aggregates boundary cues, neighboring context, and the chord features at frame $t$.

The final decoder of BACHI is a single-layer transformer decoder block consisting of a self-attention module and a cross-attention module. Its output is a framewise chord-element sequence $\mathbf{X}_t \in R^{3\times d\_model}$. During training, we adopt the masked transformer paradigm, where each decoder input $\mathbf{X}_t^m$ is a randomly masked version of $\mathbf{X}_t$. The model is optimized to fill the masked elements. The decoder leverages self-attention over $\mathbf{X}^m_t$ and cross-attention over $\mathcal{C}_t$, thereby integrating information from both existing chord elements ($r,q,b$), and the local context encoded in $\mathcal{C}_t$:
\begin{align} 
\mathbf{X}_t &\leftarrow \mathrm{FeedForward}(\mathrm{CA}(\mathrm{SA}(\mathbf{X}_t^m), \mathcal{C}_t, \mathcal{C}_t)) 
\end{align}
where $\mathrm{CA(Q,K,V)}$ denotes the cross attention computation.
Then the logits \(\ell^s_t\) are obtained for each chord element via separate classification heads. Note that the decoder \textbf{does not} auto-regressively predicts the output, as all masked fields are predicted simultaneously at one time. 

\begin{table*}[t]
  \centering
  \resizebox{0.9\textwidth}{!}{
  \begin{tabularx}{\textwidth}{
  >{\raggedright\arraybackslash}p{4.0cm} | 
  *{8}{>{\centering\arraybackslash}X}
}
    \toprule
      & \multicolumn{8}{c}{\textbf{Accuracy} \;(\accunit)} \\
    \cmidrule(lr){2-9}
      \textbf{Model / Approach}  & \multicolumn{4}{c}{\textbf{Classical Corpus}} 
      & \multicolumn{4}{c}{\textbf{Pop909-CL}} \\
    \cmidrule(lr){2-5}\cmidrule(lr){6-9}
      & \textbf{Root} & \textbf{Quality} & \textbf{Bass} & \textbf{Full}
      & \textbf{Root} & \textbf{Quality} & \textbf{Bass} & \textbf{Full} \\
    \midrule

    Rule-based~\cite{rule-based}
      & 54.6 & 45.8 & 50.5 & 28.4
      & 85.9 & 69.7 & 85.8 & 65.0 \\
    AugmentedNet~\cite{augmentednet}
      & 73.9 & 74.2 & 72.3 & 57.2
      & 88.6 & 84.5 & 90.5 & 78.7 \\
    ChordGNN~\cite{chordgnn}
      & 73.0 & 73.7 & 71.0 & 58.5
      & 80.7 & 82.0 & 82.7 & 71.6 \\ 
    Harmony Transformer v2~\cite{htv2}
      & 76.1 & 76.8 & 75.2 & 62.1
      & \textbf{90.5} & \textbf{86.9} & \textbf{92.1} & 82.2 \\
    BACHI (ours)
      & \textbf{77.8} & \textbf{79.0} &\textbf{77.0}& \textbf{68.1}
      & 89.6 & 86.8 & 91.3 & \textbf{82.4} \\
    \bottomrule
  \end{tabularx}}
  \caption{Model performance on classical Corpus (DCML and WiR) and POP909-CL. Accuracies are reported \accunit.}
  \label{tab:comparisons}
  \vspace{-0.4cm}
\end{table*}

At inference, we follow the steps below to obtain the final chord prediction:
\begin{enumerate}[leftmargin=*, itemsep=1pt, topsep=2pt]
    \item  initialize $\mathbf{X}_t^m$ to all \texttt{[MASK]} and send to the decoder.
    \item Compute confidences of the output \(c^s_t=\max(\mathrm{softmax}(\ell^s_t))\) for each unfilled \(s\in\{q,r,b\}\).
    \item Commit the highest-confidence element prediction.
    \item Repeat until all components are filled (three iterations in total).
\end{enumerate}
This results in a simple, order-agnostic procedure that adapts the prediction order to the data and mirrors human ear-training practices, by first identifying the most salient element and then progressively resolving the remaining ones.

We train the decoder jointly with the other components as an end-to-end model. Since the input to the beginning transformer encoder has the different size of the input to the final decoder, in code implementation, we conduct the reshaping along the dimensions of batch size and contextual length to support the end-to-end training.


\section{Experiments}\label{sec:experiments}

\subsection{Dataset and Training Setup}
\label{subsec:classical_corpus}

\noindent\textbf{Classical Music} \; We construct a classical corpus by combining When-in-Rome (WiR)~\cite{when-in-rome} and DCML~\cite{dcml} functional-harmony repositories and make the de-duplications. Since both datasets focus on a harmonic analysis task instead of chord recognition, we convert their annotations into the absolute chord labels via \textit{music21}~\cite{music21} package and a self-written conversion script of chord quality. 

\vspace{0.2cm}

\noindent\textbf{POP909-CL} \; The original POP909 dataset~\cite{pop909} consists of piano arrangements of 909 Chinese pop songs in MIDI format. Although it provides the extracted beat, chord, and key annotations, many of these contain errors due to the limitations of the rule-based extraction algorithms. In addition, the tempo varies within each sample, preventing direct conversion into score-aligned symbolic data with a fixed tempo.

To address these limitations, we introduce POP909-CL, an enhanceedd version of POP909 with \textbf{C}orrect \textbf{L}abels of chords, beats, keys, and time signatures across all 909 tracks. We recruit professional musicians to refine the annotations: starting from the rule-based extraction labels, they carefully reviewed each track and corrected erroneous labels. During this process, they also provided comments on special cases (e.g., the presence of weak attack bars), which we manually resolved to ensure consistency and accuracy.

\begin{figure}[t]
    \centering
    \includegraphics[width=\linewidth]{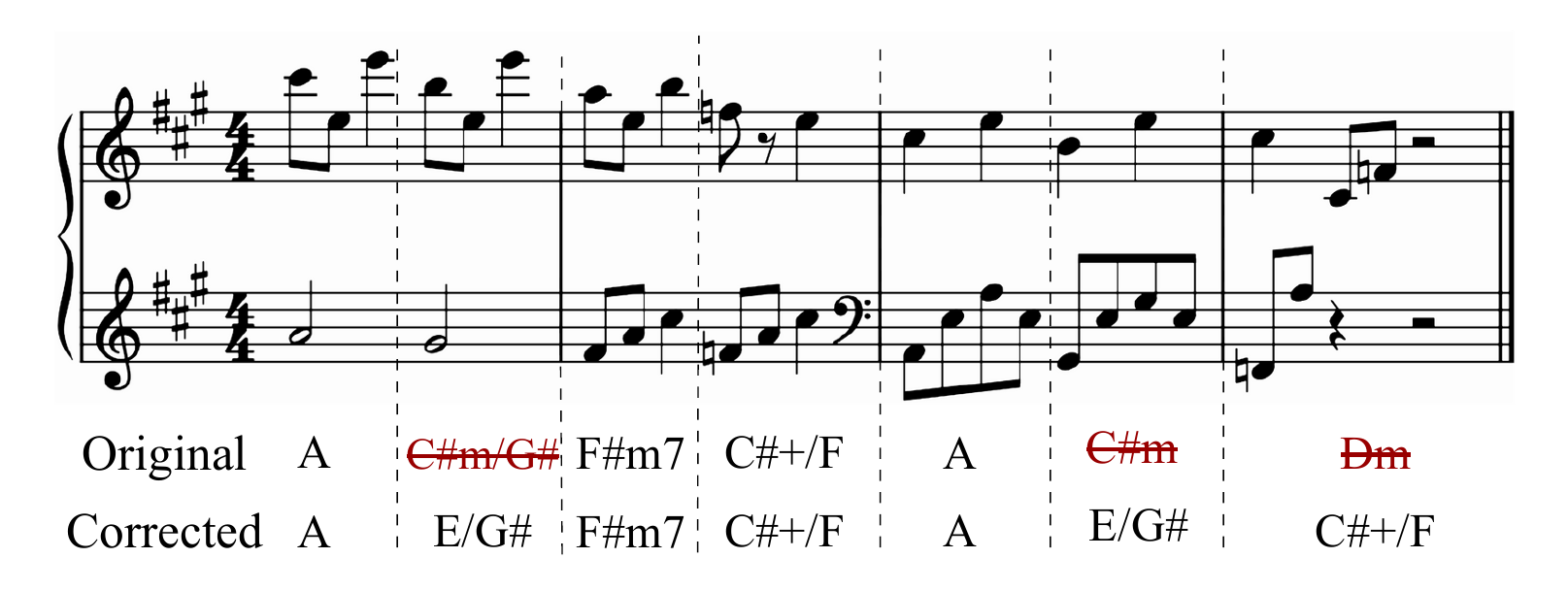}
    \vspace{-0.7cm}
    \caption{The chord label comparison between the rule-based extraction (original in POP909) to human corrected ones (in POP909-CL).}
    \label{fig:human_label}
    \vspace{-0.5cm}
\end{figure}

A statistical comparison between POP909 and POP909-CL is shown below: 40.6\% of start beats in POP909 are misaligned, 14.2\% of key signature changes are missing, and 2.6\% of time signatures are incorrect. And in POP909-CL, we correct all of them. Figure~\ref{fig:human_label} provides an example of human-corrected labels versus algorithmic extraction. These refinements establish POP909-CL as a reliable pop music dataset with corrected annotations for both analytical and generative tasks. We will release POP909-CL to the music community.
\vspace{0.2cm}

\noindent\textbf{Training Specification} \; We collect 1500 unique classical and 909 pop music pieces with accurate chord labels for the training. Since we collect the correct beat labels in POP909-CL, we aligned all POP909 samples with a fixed tempo to become a music score version for ACR training. 
We split each dataset using 9:1 train-test splits, apply 12-key augmentation for the training sets, and only use the canonical version in the test set. We do not train a single with both classical and pop music data but two separate models for each, because the data distribution and the chord pattern are largely deviated through these two genres as described in Section~\ref{sec:intro}.

\vspace{0.2cm}
\noindent\textbf{Training Setting} \;\label{subsec:hparams}
We optimize with AdamW~\cite{adamw} (\(\beta_1{=}0.9,\beta_2{=}0.98\), \(\mathrm{eps}{=}10^{-9}\)), and apply linear warm-up for 4000 steps (classical) and 2000 steps (POP909-CL), followed by cosine decay on the learning rate range (1e-5, 1e-4). We employ the mixed-precision training in bfloat16 and set the maximum gradient clip norm to 2.0.

\subsection{Evaluation, Ablations and Baselines}\label{subsec:metrics}
We evaluate predictions using macro-accuracy over chord elements (root, quality, and bass) as well as overall chord accuracy. As baselines, we select AugmentedNet~\cite{augmentednet}, ChordGNN~\cite{chordgnn}, and Harmony Transformer v2~\cite{htv2}. We retrain all three models on the same classical and pop datasets used for BACHI to ensure fair comparison. For Harmony Transformer v2, we modify the output format from a single target to three targets ($r,q,b$) to match our setup. In addition, we include the training-free rule-based method~\cite{rule-based} as a lower-bound anchor, extracting root, quality, and bass from its chord predictions, since the chord label is its only output.

We also conduct ablation studies on four variants: (1) BACHI without boundary detection and iterative decoding, directly projecting encoder outputs to chord labels; (2) BACHI without iterative decoding; (3) BACHI conditioned on additional key detection, implemented by training an auxiliary MLP for key prediction and embedding its output via FiLM together with boundary information; and (4) the full BACHI model. Due to page limitations, we report all accuracy comparison and ablation results on classical datasets in the main text, with POP909-CL results and full MIREX-style comparisons provided in the demo website.\footnote{\url{https://andyweasley2004.github.io/BACHI/}}. 

\section{Results}

\subsection{Comprehensive Performance}
Table~\ref{tab:comparisons} presents results across both classical and pop music evaluation sets. For classical music, BACHI achieves the highest scores across all metrics, with a notable improvement in full chord accuracy (68\%) over prior baselines. Nevertheless, chord recognition in classical music remains highly challenging, as all models perform the full chord accuracy below 70\%. This reflects the greater harmonic complexity and stylistic biases across composers and periods, highlighting the need for future work to better address these challenges. 

In contrast, most models achieve above 75\% full chord accuracy on pop music, and the gaps among them are smaller. BACHI attains the best results in full chord and chord quality accuracy, and ranks second in root and bass accuracy, following Harmony Transformer v2~\cite{htv2}. We also observe that the rule-based method used in the original POP909 dataset achieves only 65\% chord accuracy. This underscores the value of POP909-CL, which corrects approximately 35\% of chord label errors in POP909.

\begin{figure}[h]
    \centering
    \vspace{-0.8cm}
    \includegraphics[width=1\linewidth]{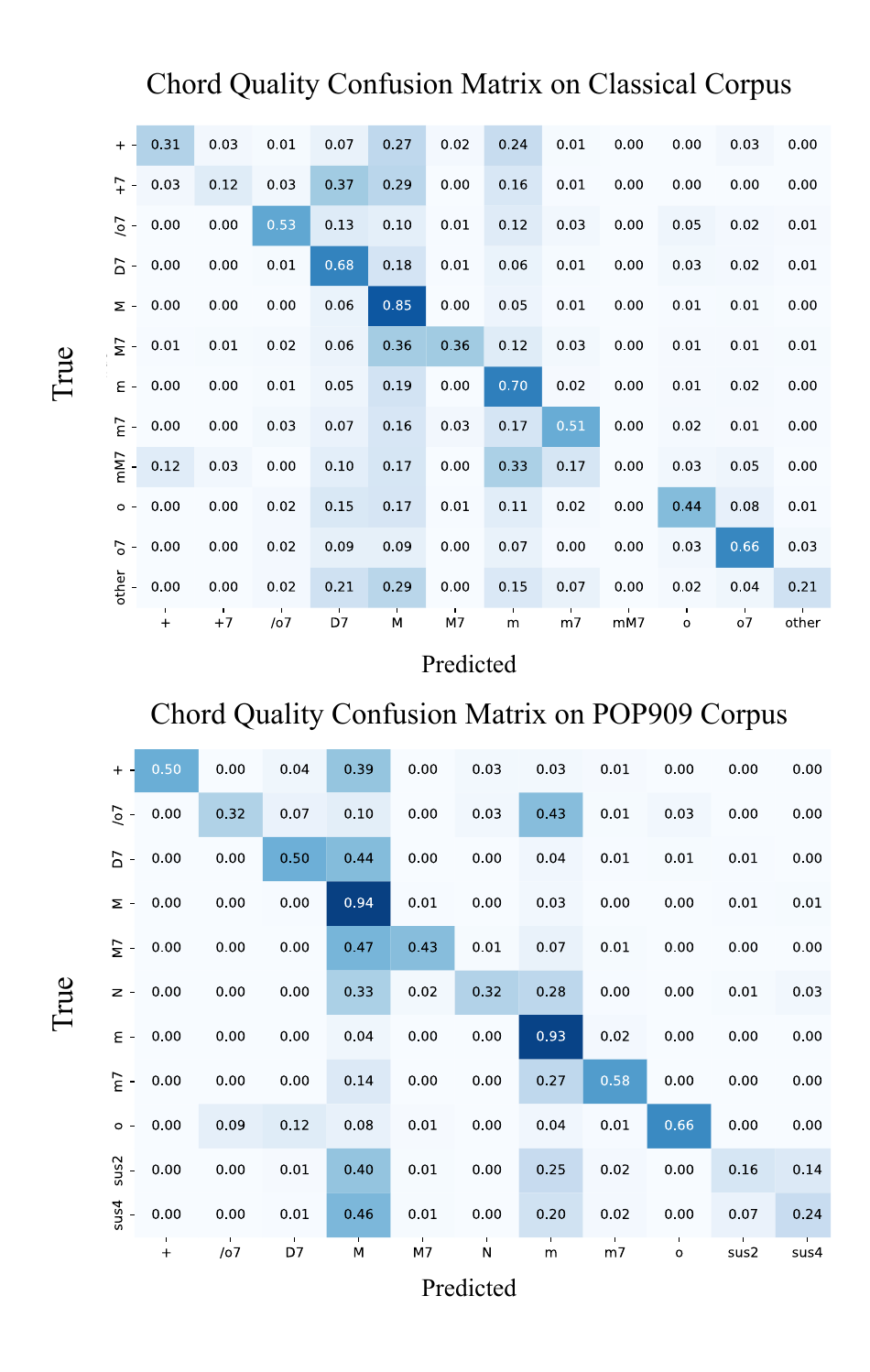}
    \vspace{-1.3cm}
    \caption{Confusion matrices on chord quality in classical corpus and POP909-CL evaluation sets.}
    \vspace{-0.6cm}
    \label{fig:confusion}
\end{figure}

Confusion matrices in Figure~\ref{fig:confusion} further illustrate the difference between pop and classical music chord patterns. On POP909-CL, most errors involve confusion between closely related qualities (e.g., major vs. minor), while in classical music, misclassifications are distributed more broadly across qualities. These patterns confirm that pop harmony is relatively predictable and concentrated, whereas classical harmony exhibits greater variability and annotation ambiguity.

Overall, these results demonstrate both the strong performance of BACHI on classical and pop music and the contribution of POP909-CL as a reliable resource for symbolic ACR.
\begin{table}[t]
  \smalltable  
  \centering
  \begin{tabularx}{\linewidth}{
    >{\raggedright\arraybackslash}X
    | c c c | c
  }
    \toprule
    \textbf{Model Design} 
      & \multicolumn{4}{c}{\textbf{Accuracy} \;(\accunit)} \\ 
    \cmidrule(lr){2-5}
      & \textbf{Root} & \textbf{Quality} & \textbf{Bass} & \textbf{Full Chord} \\
    \midrule
    BACHI w/o. \textbf{BD} and \textbf{ID}
      & \textbf{78.0} & 78.9 & \textbf{77.3} & 66.1 \\
    
    BACHI w/o. \textbf{ID}
      & 77.8 & 78.8 & 76.8 & 65.6 \\
    BACHI w/. key detection
      & 77.4 & 78.6 & 76.4 & 67.6 \\
    BACHI
      & 77.8 & \textbf{79.0} & 77.0 & \textbf{68.1} \\
    \bottomrule
  \end{tabularx}
  \vspace{-4mm}
  \caption{Ablation study on our BACHI variants. The accuracy metrics are reported \accunit.}
  \label{tab:ablation-chord}
  \vspace{-6mm}
\end{table}
Overall, our approach is particularly effective on challenging classical repertoire while remaining competitive on popular music, demonstrating robustness across diverse harmonic contexts.

\subsection{Ablation Study}
Table~\ref{tab:ablation-chord} presents ablation results of BACHI variants on the classical corpus, where \textbf{BD} denotes boundary detection and \textbf{ID} denotes iterative decoding. Without BD and ID, BACHI reduces to a basic transformer encoder model, achieving reasonable individual component performance (78.0\% root, 78.9\% quality, 77.3\% bass) but lower full chord accuracy (66.8\%). 

Adding BD primarily helps by reducing jitter (in our results, this gain is most evident on pop). However, the absence of iterative decoding still limits performance, indicating that iterative decoding across chord elements contributes significantly to overall accuracy. Interestingly, adding key detection as an additional condition slightly decreases full-chord accuracy compared to the full BACHI, likely due to errors in key prediction propagating to chord recognition.


These findings demonstrate that the combination of boundary detection and iterative decoding substantially enhances chord recognition, while adding extra conditions, such as key, does not necessarily improve performance due to potential error accumulation in auxiliary detection tasks.

\section{Discussion}
Our confidence-guided decoding reveals striking repertoire-specific patterns that validate our human ear-training practices. In classical pieces, the model tends to predict quality first (with the ratio 40.8\%). The most frequent prediction chain is quality$\rightarrow$root$\rightarrow$bass (33.2\%), matching analysis that infers chord type from voice-leading. In POP909-CL, the model tends to predict bass first (66.9\%), as the most frequent chain is bass$\rightarrow$root$\rightarrow$quality (56.4\%). This is consistent with bass-led cues in pop. These genre-variant orders indicate that the model internalizes musician-like heuristics, supporting our hypothesis that human-mimicking decision paths benefit symbolic ACR and improve upon fixed-order decoding.

\section{Conclusion}
In this research, we presented BACHI, a boundary-aware transformer for symbolic ACR, combining patch embedding, FiLM-based boundary conditioning, and confidence-guided masked-filling decoding. Our model achieves substantial improvements over prior baselines, and we contribute new human-annotated chord labels for the POP909-CL dataset to support evaluation and generation tasks. Analysis further shows that confidence-ordered decoding adapts to genre-specific patterns, with classical music favoring quality-first prediction and popular music favoring bass-first prediction, remarking the value of flexible decoding strategies under various uncertainty levels in multiple tasks.  

Our boundary-aware framework and POP909-CL annotations provide a foundation for advancing generative systems, enabling more harmonically coherent and structurally aware generation under chord conditions. Beyond applications in generation, learned confidence patterns also offer new insights into music theory by highlighting repertoire-specific harmonic tendencies. We have released our code, trained models, and POP909-CL annotations, which can be accessed on our demo page \footnote{\url{https://andyweasley2004.github.io/BACHI/}}.

\vfill\pagebreak

\bibliographystyle{IEEEbib}
\bibliography{reference}

\end{document}